\documentclass[12pt]{article}
\usepackage{amssymb,amsfonts,amsmath}
\begin{document}
 \hfill\break
\begin{center}\vspace{2.0cm}{\bf Hamilton-Jacobi Treatment of The Scalar
Field Coupled to Two Fermions}\\

 \vspace{0.5cm}{\bf Walaa I. Eshraim}\\
New York University Abu Dhabi, Saadiyat Island, P.O. Box 129188, Abu Dhabi, U.A.E.\\

\vspace{1.3cm}{\bf
 Abstract}\end{center} \vspace{0.4cm}
$~~~$The constrained filed system, the scalar field coupled to two flavours of fermions through
Yukawa couplings, is treated by using the
Hamilton-Jacobi approach. The equations of motion are obtained as
total differential equations in many variables. The equations of
motion are in exact agreement with those equations obtained using
Dirac's method. Due to the Hamilton-Jacobi quantization, the path integral 
of the scalar field coupled to two fermions is obtained directly
as an integration over the canonical phase space without using any gauge fixing condition.
$$$$
keywords: Field Theory; Gauge Fields; Hamilton-Jacobi Formulation; Singular Lagrangian, Path Integral Quantization of Constrained Systems.

 \vfill\eject

\section{Introduction}
\indent Investigation the Hamiltonian treatment
of constrained systems was initiated by Dirac [1], which is the most common method. Considering the primary constraints first is the main feature
of his method. All constraints are obtained using consistency conditions. Besides, he
showed that the number of degrees of freedom of the dynamical
system can be reduced. Hence, the equations of motion of constrained system are obtained in terms of arbitrary parameters.\\
\indent The canonical method, or G\"{u}ler's method, developed
Hamilton-Jacobi approach [2-3] to investigate constrained systems. In Refs. [4, 5], the equivalent Lagrangian method is used to obtain the set of Hamilton-Jacobi partial differential equations
(HJPDE). In the Hamilton-Jacobi treatment of constrained systems [6-9]: (i) the distinction between first- and second-class constraints is unnecessary (ii) the equations of
motion are obtained as total differential equations in many variables (iii) the investigation of the integrability conditions is required (iv) that gauge fixing is not necessary. \\
\indent Quantization of the constrained Hamiltonian systems can be achieved
by using several methods which are operator methods [1, 10], by path integral quantization [11-13], and based on the canonical method [14], which called the Hamilton-Jacobi quantization [15-20].\\
\indent We investigate in the present work the scalar field coupled to two
flavours of fermions through Yukawa couplings by using both Hamilton-Jacobi formulation and Dirac's method. Note that, Yukawa couplings are named after Hideki
Yukawa interacted a scalar field $\varphi$ with a
Dirac field $\psi$ of the type, which can be used to describe the
strong nuclear force between nucleons, mediated by pions, and used
in the Standard Model to describe the coupling between the Higgs
field and massless quark and electron fields. Through spontaneous
symmetry breaking, the fermions acquire a mass proportional to the
vacuum expectation value of the Higgs field. The path integral quantization of the  scalar field coupled to two flavours of fermions will be also quantize by using the Hamilton-Jacobi quantization.\\
\indent In this paper both methods, Dirac's and G\"{u}ler's are
used to tackle the system of the scalar field coupled to two
flavours of fermions through Yukawa couplings. This paper is
arranged as follows: Hamilton-Jacobi approach is presented in Sec.2. Dirac's method is used in Sec.3. 
Hamilton-Jacobi method is applied in sec.4 and the quantization of the system in Sec.5. The paper closes with a conclusion in Sec. 6.


\section{Hamilton-Jacobi approach}

\indent The Hamilton-Jacobi approach can be introduced
as follows:\\
If the rank of the Hess matrix is:
\begin{equation}\label{1}
A_{ij}=\frac{\partial^{2}L}{{\partial{\dot{q}}_{i}}\>{\partial{\dot{q}}_{j}}}\>,
\end{equation}
is $(n-r),  r<n$, then the standard definition of a linear momenta
is:
\begin{align}
p_{a}&= \frac{\partial{L}}{\partial{\dot{q}}_{a}},\qquad {a =
1,2,\ldots,n-r}, \label{2}\\
 p_{\mu}&=\frac{\partial{L}}{{\partial{\dot{q}}_{\mu}}},\qquad {\mu =
n-r+1,\ldots,n},\label{3}
\end{align}
enables us to solve Eq.(2) for ${\dot{q}}_{a}$ as
\begin{equation}\label{4}
{\dot{q}}_{b}=\omega_{b}(q_{i},{{\dot{q}}_{\mu}},p_{a}).
\end{equation}
Substituting Eq.(4), into Eq.(3), we obtain the constraints as
\begin{equation}\label{5}
H'_{\mu}= p_{\mu}+H_{\mu}(\tau,q_{i},p_{a})=0,
\end{equation}
where
\begin{equation}\label{6}
H_{\mu}=
-\frac{\partial{L}}{\partial{\dot{q}}_{\mu}}\bigg|_{{\dot{q}}_{a}
\equiv\omega_a}.
\end{equation}
The usual Hamiltonian $H_{0}$ is defined as:
\begin{equation}\label{7}
H_{0}=-L+ p_{a}{{\dot{\omega}}_a}-{{\dot{q}}_{\mu}}H_{\mu}.
\end{equation}
Like functions $H_{\mu}$, the function $H_{0}$ is not an explicit
function of the velocities ${\dot{q}}_{\nu}$. Therefore, the
Hamilton-Jacobi function $S(\tau,q_{i})$ should satisfy the
following set of Hamilton-Jacobi partial differential equations
(HJPDE) simultaneously for an extremum of the function:
\begin{equation}\label{8}
H'_{\alpha}\bigg(t_{\beta},~q_{\alpha},~P_{i}=\frac{\partial
S}{\partial q_{i}} ,~P_{0}=\frac{\partial S}{\partial
t_{0}}\bigg)=0,
\end{equation}
where\\
$\qquad {\alpha,\beta =0,n-r+1,\ldots,n};\qquad {a =
1,2,\ldots,n-r}$,and
\begin{equation}\label{9}
H'_{\alpha}=p_{\alpha}+H_{\alpha}.
\end{equation}
The canonical equations of motion are given as total differential
equations in variables $t_{\beta}$,
\begin{align}
dq_{p}&=\,\frac{\partial H'_{\alpha}}{\partial p_{p}}\,
dt_{\alpha},\quad {p= 0,1,\ldots,n};\quad {\alpha= 0,n-r+1,\ldots,n},\label{10}\\
dp_{a}&=-\frac{\partial H'_{\alpha}}{\partial q_{a}}\,
dt_{\alpha},\;\qquad a=1,\ldots,n-r,\label{11}\\
dp_{\mu}&=-\frac{\partial H'_{\alpha}}{\partial
q_{\mu}}\,dt_{\alpha}, \,\;\quad \quad \alpha =
0,n-r+1,\ldots,n,\label{12}
\end{align}
\begin{equation}\label{13}
dZ= \left(-H_{\alpha}+p_{a}\frac{\partial H'_{\alpha}}{\partial
p_{\alpha}}\,dt_{\alpha}\right),
\end{equation}
where
\begin{equation}\label{14}
Z\equiv S(t_{\alpha},q_{a}),
\end{equation}
being the action. Thus, the analysis of a constrained system is
reduced to solve equations (10-12) with constraints
\begin{equation}\label{15}
H'_{\alpha}(t_{\beta},~q_{a},~P_{i})=0,\qquad {\alpha,\beta
=0,n-r+1,\ldots,n}.
\end{equation}
Since the equations above are total differential equations,
integrability conditions should be checked. These equations of
motion are integrable [3-5,8] if and only if the variations of
$H'_{\alpha}$ vanish identically as:
\begin{equation}\label{16}
dH'_{\alpha}=0.
\end{equation}
If they do not vanish identically, then we consider them as new
constraints. This procedure is repeated until a complete system is
obtained.\\

\subsection{Hamilton-Jacobi Quantization}

Path integral
formulation of the constrained systems is studied in Refs. [14-20]. One has to consider a singular Lagrangian, as seen in previous section, for computing the Hamilton-Jacobi path integral. The canonical Hamiltonian $H_0$ defined in Eq. (6), and the set of
HJPDE is expressed in Eq. (7). As we define
\begin{equation}
p_{\beta}=\frac{\partial S[q_a;x_\alpha]}{\partial x_\beta}\,,
\end{equation}
 and
 \begin{equation}
p_a=\frac{\partial S[q_a;x_\alpha]}{\partial q_a}\,,
\end{equation}
with $x_0=t$
and $S$ being the action. The total differential equations given
in (10-13) are integrable if (16) are hold \cite{21}.
If conditions (16) are not satisfied identically, one
considers them as new constraints and again
consider their variations.\\
\indent Thus, repeating this procedure one may obtain a set of
constraints such that all variations vanish. Simultaneous
solutions of canonical equations with all these constraints
provide to obtain the set of canonical phase space coordinates
$(q_a,p_a)$ as functions of $t_\alpha$, besides the canonical
action integral is obtained in terms of the canonical coordinates.
$H'_\alpha$ can be interpreted as the infinitesimal generator of
canonical transformations given by parameters $t_\alpha$. In this
case path integral representation may be written as 
\begin{multline}\label{HJQ}
\left<Out\mid S\mid In\right> = \int \prod^{n-p}_{a=1}dq^a dp^a
exp \left\{\int_{t_\alpha}^{t'_\alpha} \left(-H_\alpha + p_\alpha
\frac{\partial H'_\alpha}{\partial p_\alpha}\right)\, dt_\alpha
\right\},\\\qquad a=1,\ldots,n-p,\qquad \alpha=0,n-p+1,\ldots,n\,.
\end{multline}
\indent In fact, this path integral is an integration over the
canonical phase-space coordinates $(q^a, p^a)$.

\section{Dirac's method}

In this section, we treat the constrained field system by Dirac's method [22].\\

Consider one loop order the self-energy for the scalar field
$\varphi$, with a mass $m$, coupled to two flavours of fermions,
with masses $m{_1}$ and $m{_2}$, due to Yukawa couplings 

\begin{multline}\label{20}
L={\frac{1}{2}}(\partial_{\mu}\varphi)^2-{\frac{1}{2}}{m^2}{\varphi^2}-{\frac{1}{6}}\lambda
{\varphi^3}+\sum_{i}{{\overline{\psi}}_{(i)}}(i{\gamma^\mu}\partial_\mu-m_i){\psi_{(i)}}\\
-g\varphi({{\overline{\psi}}_{(1)}}{\psi_{(2)}}+{{\overline{\psi}}_{(2)}}{\psi_{(1)}}),
\qquad{\mu =0,1,2,3},
\end{multline}
where $\lambda$ is parameter and $g$ constant, $\varphi$,
$\psi_{(i)}$, and ${\overline\psi}_{(i)}$ are odd ones. We are
adopting the Minkowski metric
$\eta_{\mu\nu}=diag(+1,-1,-1,-1)$.\\
\indent The Lagrangian function (20) is singular, since the rank
of the Hess matrix
\begin{equation}\label{18}
A_{ij}=\dfrac{\partial^{2}L}{{\partial{\dot{q}}_{i}}\>
{\partial{\dot{q}}_{j}}},
\end{equation}is one.\\
The generalized momenta (2,3)
\begin{equation}\label{19}
p_{\varphi} =\frac{\partial
L}{\partial\dot{\varphi}}={\partial^0}\varphi,
\end{equation}
\begin{equation}\label{20}
p_{(i)} =\frac{\partial L}{\partial{\dot{\psi}}_{(i)}}=
i{{\overline{\psi}}_{(i)}}{\gamma^0}= -H_{(i)},\qquad {i =1,2},
\end{equation}
\begin{equation}\label{21}
\overline{p}_{(i)}=\frac{\partial
L}{\partial{\dot{\overline\psi}}_{(i)}}=0=-\overline{H}_{(i)}.
\end{equation}
Where we must call attention to the necessity of being careful
with the spinor indexes. Considering, as usual $\psi_{(i)}$ as a
column vector and ${\overline\psi}_{(i)}$ as a row vector implies
that $p_{(i)}$ will be a row vector while ${\overline p}_{(i)}$
will be a column vector. \\
 Since the rank of the Hess matrix is one, one may
solve (22) for ${\partial^0}\varphi$ as:
\begin{equation}\label{22}
{\partial^0}\varphi=p_{\varphi}\equiv\omega.
\end{equation}
The usual Hamiltonian $H_0$ is given as:
\begin{equation}\label{23}
H_0 = -L+\omega
p_{\varphi}+{{\partial_0}\psi_{(i)}}\>p_{(i)}\>{\bigg|_{{p_{(i)}=-H_{(i)}}}}+{
{{\partial_0}{\overline{\psi}_{(i)}}}\>{\overline
p}_{(i)}}\>\bigg|_ {{\overline p}_{(i)}= -{\overline H}_{(i)}},
\end{equation}
or
\begin{multline}\label{24}
H_0 =
{\frac{1}{2}}({p^2}_{\varphi}-{\partial_{a}\varphi}{\partial^a}\varphi)+{\frac{1}{2}}{m^2}{\varphi^2}+{\frac{1}{6}}\lambda
{\varphi^3}-{{\overline{\psi}}_{(i)}}(i{\gamma^a}\partial_{a}-m_i){\psi_{(i)}}
\\+g\varphi({{\overline{\psi}}_{(1)}}{\psi_{(2)}}+{{\overline{\psi}}_{(2)}}{\psi_{(1)}}),\qquad {a=1,2,3.}
\end{multline}
Eqs. (23), and (24) lead to the primary constraints
\begin{equation}\label{25}
H'_{(i)}= p_{(i)}+H_{(i)}=
p_{(i)}-i\>{{\overline{\psi}}_{(i)}}\,{\gamma^0}=0,
\end{equation}
and
\begin{equation}\label{26}
\overline H'_{(i)}= {\overline p}_{(i)}+{\overline
H}_{(i)}={\overline p}_{(i)}=0,
\end{equation}
respectively. These constraints lead to the total Hamiltonian
\begin{equation}\label{27}
H_T=H_0+{\lambda_{(i)}}H'_{(i)}+{{\overline\lambda}_{(i)}}\overline
H'_{(i)},
\end{equation}
or
\begin{multline}\label{28}
H_T={\frac{1}{2}}({p^2}_{\varphi}-{\partial_{a}\varphi}{\partial^a}\varphi)+
{\frac{1}{2}}{m^2}{\varphi^2}+{\frac{1}{6}}\lambda{\varphi^3}
-{{\overline{\psi}}_{(i)}}(i{\gamma^a}\partial_{a}-m_i){\psi_{(i)}}\\
+g\varphi({{\overline{\psi}}_{(1)}}{\psi_{(2)}}+{{\overline{\psi}}_{(2)}}{\psi_{(1)}})
+{\lambda_{(i)}}(p_{(i)}-i{{\overline{\psi}}_{(i)}}{\gamma^0})+{{\overline\lambda}_{(i)}}{\overline
p}_{(i)}.
\end{multline}
According to Dirac's method, the time derivative of the primary
constraints should be zero, that is
\begin{equation}\label{29}
\dot{H'}_{(1)}=\{H'_{(1)},H_T\}=
{\overline\psi}_{(1)}(i\overleftarrow{\partial_a}\gamma^a+m_{1})+g\,\varphi{\overline\psi}_{(2)}-i{\overline{\lambda}_{(1)}}\gamma^0\approx0,
\end{equation}
\begin{equation}\label{30}
\dot{H'}_{(2)}=\{H'_{(2)},H_T\}={\overline\psi}_{(2)}(i\overleftarrow{\partial_a}\gamma^a
+m_{2})+g\,\varphi\,{\overline\psi}_{(1)}-i{\overline{\lambda}_{(2)}}\gamma^0\approx0,
\end{equation}
\begin{equation}\label{31}
\dot{\overline H'}_{(1)}=\{\overline
H'_{(1)},H_T\}=-(i{\gamma^a}{\partial_a}-m_{1}){\psi_{(1)}}
+g\varphi\,{\psi_{(2)}}-i{\gamma^0}\>{\lambda_{(1)}}\approx0,
\end{equation}
\begin{equation}\label{32}
\dot{\overline H'}_{(2)}=\{{\overline H'}_{(2)},H_T\}=
-(i{\gamma^a}{\partial_a}-m_{2}){\psi_{(2)}}
+g\varphi\,{\psi_{(1)}}-i{\gamma^0}\>{\lambda_{(2)}}\approx0.
\end{equation}
\\
Eqs. (32-35) fix the multipliers $\overline\lambda_{(1)}$,
$\overline\lambda_{(2)}$, $\lambda_{(1)}$, and $\lambda_{(2)}$
respectively as
\begin{equation}\label{33}
i{\overline{\lambda}_{(1)}}\gamma^0={\overline\psi}_{(1)}(i\overleftarrow{\partial_a}\gamma^a
+m_{1})+g\,\varphi\,{\overline\psi}_{(2)},
\end{equation}

\begin{equation}\label{34}
i{\overline{\lambda}_{(2)}}\gamma^0={\overline\psi}_{(2)}(i\overleftarrow{\partial_a}\gamma^a
+m_{2})+g\,\varphi\,{\overline\psi}_{(1)},
\end{equation}

\begin{equation}\label{35}
i{\gamma^0}\>{\lambda_{(1)}}=-(i{\gamma^a}{\partial_a}-m_{1}){\psi_{(1)}}
+g\varphi\,{\psi_{(2)}},
\end{equation}

\begin{equation}\label{36}
i{\gamma^0}\>{\lambda_{(2)}}=-(i{\gamma^a}{\partial_a}-m_{2}){\psi_{(2)}}
+g\varphi\,{\psi_{(1)}}.
\end{equation}
\\
Multiplying Eqs. (36) and (37) from the right and Eqs. (38) and (39)
from the left by $-i\gamma^0$, we obtain:
\begin{equation}\label{40}
{\overline{\lambda}_{(1)}}={\overline\psi}_{(1)}\,(\overleftarrow{\partial_a}\gamma^a
-im_{1})\,\gamma^0-ig\,\varphi\,{\overline\psi}_{(2)}\gamma^0,
\end{equation}

\begin{equation}\label{38}
{\overline{\lambda}_{(2)}}={\overline\psi}_{(2)}(\overleftarrow{\partial_a}\gamma^a
-im_{2})\,\gamma^0-ig\,\varphi\,{\overline\psi}_{(1)}\,\gamma^0,
\end{equation}

\begin{equation}\label{39}
\lambda_{(1)}=-\gamma^0({\gamma^a}{\partial_a}+im_{1}){\psi_{(1)}}
-i\,g\varphi\,\gamma^0\>{\psi_{(2)}},
\end{equation}

\begin{equation}\label{40}
\lambda_{(2)}=-\gamma^0({\gamma^a}{\partial_a}+im_{2}){\psi_{(2)}}
-ig\varphi\,\gamma^0\,{\psi_{(1)}}.
\end{equation}
\\
There are no secondary constraints. Taking suitable linear combinations of constraints,
 one has to find all numbers of second-class ones, there are\\
\begin{equation}\label{41}
\Phi_{i}=H'_{(i)}=
p_{(i)}-i{{\overline{\psi}}_{(i)}}\,{\gamma^0},\qquad {i =1,2},
\end{equation}
\begin{equation}\label{42}
\Phi_{3}={\overline H'_{(1)}}=\overline p_{(1)},
\end{equation}
and
\begin{equation}\label{43}
\Phi_{4}={\overline H'_{(2)}}=\overline p_{(2)}.
\end{equation}
The equations of motion are read as
\begin{equation}\label{44}
\dot{\varphi}=\{\varphi,H_T\}=p_{\varphi},
\end{equation}

\begin{equation}\label{45}
{\dot{\psi}}_{(i)}=\{\psi_{(i)},H_T\}=\lambda_{(i)},
\end{equation}

\begin{equation}\label{46}
{\dot{\overline{\psi}}}_{(i)}=\{{\overline{\psi}}_{(i)},H_T\}=\overline\lambda_{(i)},
\end{equation}

\begin{equation}\label{47}
{\dot{p}}_{\varphi}=\{p_{\varphi},H_T\}=
{m^2}\varphi+{\frac{1}{2}}\lambda{\varphi^2}
+g({{\overline{\psi}}_{(1)}}{\psi_{(2)}}+{{\overline{\psi}}_{(2)}}{\psi_{(1)}}),
\end{equation}

\begin{equation}\label{48}
{\dot{p}}_{(1)}=\{p_{(1)},H_T\}={\overline\psi}_{(1)}(i\overleftarrow{\partial_{a}}\gamma^a+m_{1})
+g\varphi{\overline\psi}_{(2)},
\end{equation}

\begin{equation}\label{49}
{\dot{p}}_{(2)}=\{p_{(2)},H_T\}={\overline\psi}_{(2)}(i\overleftarrow{\partial_{a}}\gamma^a+m_{2})
+g\varphi{\overline\psi}_{(1)},
\end{equation}

\begin{equation}\label{50}
{\dot{\overline p}}_{(1)}=\{{\overline
p}_{(1)},H_T\}=-(i{\gamma^a}{\partial_a}-m_{1}){\psi_{(1)}}
+g\varphi{\psi_{(2)}}-i{\gamma^0}\lambda_{(1)},
\end{equation}

\begin{equation}\label{51}
{\dot{\overline p}}_{(2)}=\{{\overline
p}_{(2)},H_T\}=-(i{\gamma^a}{\partial_a}-m_{2}){\psi_{(2)}}
+g\varphi{\psi_{(1)}}-i{\gamma^0}\lambda_{(2)}.
\end{equation}
Differentiate Eq. (47) with respect to time, and substituting from
Eq. (50), we get:
\begin{equation}\label{52}
\ddot{\varphi}-{m^2}\varphi-{\frac{1}{2}}\lambda{\varphi^2}
-g({{\overline{\psi}}_{(1)}}{\psi_{(2)}}+{{\overline{\psi}}_{(2)}}{\psi_{(1)}})=0.
\end{equation}
Substituting from Eqs. (42) and (43) into Eqs. (48), (53) and (54),
we get:
\begin{equation}\label{56}
(i{\gamma^\mu}{\partial_\mu}-m_{1}){\psi_{(1)}}
-g\varphi\>{\psi_{(2)}}=0,
\end{equation}

\begin{equation}\label{54}
(i{\gamma^\mu}{\partial_\mu}-m_{2}){\psi_{(2)}}
-g\varphi\>{\psi_{(1)}}=0,
\end{equation}

\begin{equation}\label{55}
{\dot{\overline p}}_{(i)}=0,\qquad {i =1,2}.
\end{equation}
From Eqs. (40) and (41) into Eq. (49), we have
\begin{equation}\label{56}
{\partial_{0}}{\overline{\psi}_{(1)}}\>\>i\gamma^0-{\overline\psi}_{(1)}\,(i\overleftarrow{\partial_a}\gamma^a
+m_{1})-g\,\varphi\,{\overline\psi}_{(2)}=0,
\end{equation}

\begin{equation}\label{57}
{\partial_{0}}{\overline{\psi}_{(2)}}\>\>i\gamma^0-{\overline\psi}_{(2)}\,(i\overleftarrow{\partial_a}\gamma^a
+m_{2})-g\,\varphi\,{\overline\psi}_{(1)}=0.
\end{equation}

In the following section the same system will be discussed by
using Hamilton-Jacobi approach.

\section{Hamilton-Jacobi method}

In this section, the constrained system is going now to be tackled by using Hamilton-Jacobi treatment [23], which solve the gauge fixing problem naturally.\\

The set of (HJPDE) (8) read as
\begin{multline}\label{58}
H'_0 =p_0+H_0=p_0+
{\frac{1}{2}}({p^2}_{\varphi}-{\partial_{a}\varphi}{\partial^a}\varphi)+{\frac{1}{2}}{m^2}{\varphi^2}+{\frac{1}{6}}\lambda
{\varphi^3}-{{\overline{\psi}}_{(i)}}(i{\gamma^a}\partial_{a}-m_i){\psi_{(i)}}\\
+g\varphi({{\overline{\psi}}_{(1)}}{\psi_{(2)}}+{{\overline{\psi}}_{(2)}}{\psi_{(1)}}),
\end{multline}
\begin{equation}\label{59}
H'_{(i)}=p_{(i)}+H_{(i)}=p_{(i)}-i\>{{\overline{\psi}}_{(i)}}\,{\gamma^0}=0,
\end{equation}

\begin{equation}\label{60}
\overline H'_{(i)}=\overline p_{(i)}+\overline H_{(i)}=\overline
p_{(i)}=0.
\end{equation}

Therefore, the total differential equations for the characteristic
(10), (11) and (12) are:
\begin{equation}\label{61}
d\varphi=p_{\varphi}d\tau,
\end{equation}

\begin{equation}\label{62}
d{\psi_{(i)}}=d{\psi_{(i)}},
\end{equation}

\begin{equation}\label{63}
d{{\overline\psi}_{(i)}}=d{{\overline\psi}_{(i)}},
\end{equation}

\begin{equation}\label{64}
d{{p}_{\varphi}}=\bigg[{m^2}\varphi+{\frac{1}{2}}\lambda{\varphi^2}
+g({{\overline{\psi}}_{(1)}}{\psi_{(2)}}+{{\overline{\psi}}_{(2)}}{\psi_{(1)}})\bigg]d\tau,
\end{equation}

\begin{equation}\label{65}
d{{p}_{(1)}}=\bigg[{\overline\psi}_{(1)}(i\overleftarrow{\partial_a}\gamma^a
+m_{1})+g\,\varphi\,{\overline\psi}_{(2)}\bigg]d\tau,
\end{equation}

\begin{equation}\label{66}
d{{p}_{(2)}}=\bigg[{\overline\psi}_{(2)}(i\overleftarrow{\partial_a}\gamma^a
+m_{2})+g\,\varphi\,{\overline\psi}_{(1)}\bigg]d\tau,
\end{equation}

\begin{equation}\label{67}
d{\overline
p_{(1)}}=\bigg[-(i{\gamma^a}{\partial_a}-m_{1}){\psi_{(1)}}+g\varphi{\psi_{(2)}}\bigg]d\tau
-i{\gamma^0}d{\psi_{(1)}},
\end{equation}

\begin{equation}\label{68}
d{\overline
p_{(2)}}=\bigg[-(i{\gamma^a}{\partial_a}-m_{2}){\psi_{(2)}}+g\varphi{\psi_{(1)}}\bigg]d\tau
-i{\gamma^0}d{\psi_{(2)}}.
\end{equation}
\\
The integrability conditions $(dH'_{\alpha}=0)$ imply that the
variation of the constraints $H'_{(i)}$ and $\overline H'_{(i)}$
should be identically zero, that is
\begin{equation}\label{69}
dH'_{(i)}=dp_{(i)}-i\>d{{\overline{\psi}}_{(i)}}\,{\gamma^0}=0,
\end{equation}

\begin{equation}\label{70}
d{\overline H'_{(i)}}=d\overline p_{(i)}=0.
\end{equation}
\indent The following equations of motion:\\
 From Eq. (64), we obtain
\begin{equation}\label{71}
\dot\varphi= p_{\varphi}.
\end{equation}
 Substituting from Eqs. (68) and (69) into Eq. (72), we get

\begin{equation}\label{72}
i\partial_{0}{\overline\psi}_{(1)}\gamma^{0}-{\overline\psi}_{(1)}(i\overleftarrow{\partial_a}\gamma^a
+m_{1})-g\varphi\,{\overline\psi}_{(2)}=0,
\end{equation}

\begin{equation}\label{73}
i\partial_{0}{\overline\psi}_{(2)}\gamma^{0}-{\overline\psi}_{(2)}(i\overleftarrow{\partial_a}\gamma^a
+m_{2})-g\varphi\,{\overline\psi}_{(1)}=0.
\end{equation}
Substituting from Eqs. (70) and (71) into Eq. (73), we obtain

\begin{equation}\label{74}
(i{\gamma^\mu}{\partial_\mu}-m_{1}){\psi_{(1)}}-g\,\varphi\>{\psi_{(2)}}=0,
\end{equation}

\begin{equation}\label{75}
(i{\gamma^\mu}{\partial_\mu}-m_{2}){\psi_{(2)}}-g\,\varphi\>{\psi_{(1)}}=0.
\end{equation}
One notes that the integrability conditions are not identically
zero, they are added to the set of equations of motion.\\
From Eqs. (67-69), we get the following equations of motion:
\begin{equation}\label{76}
{\dot p}_{\varphi}={m^2}\varphi+{\frac{1}{2}}\lambda{\varphi^2}
+g({{\overline{\psi}}_{(1)}}{\psi_{(2)}}+{{\overline{\psi}}_{(2)}}{\psi_{(1)}}),
\end{equation}

\begin{equation}\label{77}
{\dot
p}_{(1)}={\overline\psi}_{(1)}(i\overleftarrow{\partial_a}\gamma^a
+m_{1})+g\,\varphi\,{\overline\psi}_{(2)},
\end{equation}

\begin{equation}\label{78}
{\dot
p}_{(2)}={\overline\psi}_{(2)}(i\overleftarrow{\partial_a}\gamma^a
+m_{2})+g\,\varphi\,{\overline\psi}_{(1)}.
\end{equation}
\\
Substituting from Eqs. (77) and (78) into (70) and (71), we get
\\
\begin{equation}\label{79}
{\dot{\overline p}}_{(i)}=0,\qquad {i =1,2}.
\end{equation}
\\
Differentiate Eq. (74) with respect to time, and making use of
(79), we obtain
\begin{equation}\label{81}
\ddot{\varphi}-{m^2}\varphi-{\frac{1}{2}}\lambda{\varphi^2}
-g({{\overline{\psi}}_{(1)}}{\psi_{(2)}}+{{\overline{\psi}}_{(2)}}{\psi_{(1)}})=0.
\end{equation}

\section{Hamilton-Jacobi quantization}

In this section, we use the Hamilton-Jacobi quantization to obtain the path integral quantization of the scalar field coupled to two flavours of fermions through Yukawa couplings.  

\indent First, one must check whether the set of equations (64-71) is
integrable or not. So, one has to consider the total variations of the
constraints. In fact\\
\begin{equation}\label{34}
dH'_{(i)}=dp_{(i)}-i\>d{{\overline{\psi}}_{(i)}}\,{\gamma^0}=0\,,
\end{equation}

\begin{equation}\label{35}
d{\overline H'_{(i)}}=d\overline p_{(i)}=0\,.
\end{equation}
The constraints (62) and (63), lead us to obtain
$d\overline{\psi}_{(i)}$ and $d\psi_{(i)}$ in terms of $dt$

\begin{equation}\label{36}
d{\overline\psi}_{(1)}i\gamma^{0}=[{\overline\psi}_{(1)}(i\overleftarrow{\partial_a}\gamma^a
+m_{1})+g\varphi\,{\overline\psi}_{(2)}]dt\,,
\end{equation}

\begin{equation}\label{37}
d{\overline\psi}_{(2)}i\gamma^{0}=[{\overline\psi}_{(2)}(i\overleftarrow{\partial_a}\gamma^a
+m_{2})+g\varphi\,{\overline\psi}_{(1)}]dt\,,
\end{equation}

\begin{equation}\label{38}
i{\gamma^0}d\psi_{(1)}=[-(i\gamma^a{\partial_a}-m_{1}){\psi_{(1)}}+g\,\varphi\>{\psi_{(2)}}]d\,,
\end{equation}
and
\begin{equation}\label{39}
i{\gamma^0}d\psi_{(2)}=[-(i\gamma^a{\partial_a}-m_{2}){\psi_{(2)}}+g\,\varphi\>{\psi_{(1)}}]dt\,.
\end{equation}
 We obtain that the set of equations (64-71)  is integrable. Making use of
(13), and  (61-63), we can write the canonical action integral as
\begin{align}\label{40}
\quad Z=\int
d^{4}x[\frac{1}{2}({p^2}_{\varphi}+{\partial_{a}\varphi}{\partial^a}\varphi)&-{\frac{1}{2}}{m^2}{\varphi^2}-{\frac{1}{6}}\lambda
{\varphi^3}+{{\overline{\psi}}_{(i)}}(i{\gamma^\mu}\partial_{\mu}-m_i){\psi_{(i)}}\nonumber \\
&
-g\varphi({{\overline{\psi}}_{(1)}}{\psi_{(2)}}+{{\overline{\psi}}_{(2)}}{\psi_{(1)}})]\,,
\end{align}
Now the path integral representation (\ref{HJQ}) is given by
\begin{align}\label{44}
\left<out|S|In\right>& =\int\prod_{i}^2\, d\varphi\,
dp_{\varphi}\,d\psi_{(i)}\,d\overline{\psi}_{(i)}\>exp\,\bigg\{i\bigg[\int
d^{4}x
\frac{1}{2}({p^2}_{\varphi}+{\partial_{a}\varphi}{\partial^a}\varphi)\nonumber \\
& -{\frac{1}{2}}{m^2}{\varphi^2}-{\frac{1}{6}}\lambda
{\varphi^3}+{{\overline{\psi}}_{(i)}}(i{\gamma^\mu}\partial_{\mu}-m_i){\psi_{(i)}} -g\varphi({{\overline{\psi}}_{(1)}}{\psi_{(2)}}+{{\overline{\psi}}_{(2)}}{\psi_{(1)}})\bigg]\bigg\}\,.
\end{align}

\section {Conclusion}
The scalar field coupled to two flavours of fermions through
Yukawa couplings is discussed as constrained system [7] by using
both Dirac's and Hamilton-Jacobi methods. In Dirac's method the
total Hamiltonian composed by adding the constraints multiplied by
Lagrange multipliers to the canonical Hamiltonian. In order to
derive the equations of motion, one needs to redefine these
unknown multipliers in an arbitrary way. However, in the
Hamilton-Jacobi approach (or G\"{u}ler's method)[2-8], there is no
need to introduce Lagrange multipliers to the canonical
Hamiltonian, then the Hamilton-Jacobi is simpler
and more economical.\\
\indent In Hamilton-Jacobi approach it is not necessary to
distinguish between first-class and second-class constraints.
Hamilton-Jacobi approach always in exact agreement with Dirac's
method. Both the consistency conditions and the integrability
conditions lead to the same constraints. Also the equations of
motion in both approaches are the same.

Hamilton-Jacobi quantization is applied to the constrained field system with finite degrees of freedom by investigating the integrability conditions without using any gauge fixing condition.

\end{document}